\documentclass[aps,preprintnumbers,nofootinbib]{revtex4}
\usepackage{amsfonts,amssymb,amsmath}            
\usepackage{amsthm}
\usepackage{subfigure}
\def\identity{\leavevmode\hbox{\small1\kern-3.8pt\normalsize1}}

\newcommand{\ket}[1]{\left |  #1 \right\rangle}

\begin{document}

\title{An Entanglement-Based Protocol For Strong Coin Tossing With
  Bias 1/4}
\date{October 22, 2006}

\author{Roger \surname{Colbeck}}
\email[]{r.a.colbeck@damtp.cam.ac.uk}
\affiliation{Centre for Quantum Computation,
             DAMTP,
	     Centre for Mathematical Sciences,
             University of Cambridge,
             Wilberforce Road,
             Cambridge CB3 0WA, UK}

\begin{abstract}
In the literature, strong coin tossing protocols based on bit
commitment have been proposed.  Here we examine a protocol that
instead tries to achieve the task by sharing entanglement
securely. The protocol uses only qubits, and has bias
$\frac{1}{4}$. This is equal to the best known bias for bit commitment
based schemes.
\end{abstract}

\maketitle

\section{Introduction}
Alice and Bob, having recently divorced, want to decide who keeps the
car \cite{Blum}.  Both now live separate lives on opposite sides of
the country, and to meet in person would be inconvenient and
traumatic.  A coin tossing protocol seeks to provide a sequence of
information exchanges that allow the decision to be fairly made.
Whether or not this is possible depends on the physical properties of
the systems used for information exchange.  Protocols which cannot
satisfy the full requirements demanded of coin tossing are given a
figure of merit depending on the maximum cheating probability they
allow.

Coin tossing comes in two flavours, weak and strong.  A {\it weak}
coin tossing protocol suffices if the parties know which outcome the
other prefers.  This is the case in the divorcees example above, where
both Alice and Bob would like to keep the car.  Suppose outcome 0
means Alice keeps the car, and 1 means Bob does.  A protocol need not
protect against Alice biasing towards 1, nor Bob towards 0, and hence
a weak coin toss protocol can be used.  In contrast, a {\it strong}
coin tossing protocol is needed when it is not known which outcome the
other party prefers.

\subsection{Previous Results} 
Coin tossing was introduced by Blum \cite{Blum} in 1981.  There, a
variant of our task was discussed in a classical setting using
computational assumptions to give security.
However, in a classical setting where unconditional
security\footnote{i.e., security based only in a belief in the laws of
physics} is sought, no protocol can offer any protection against a
cheat \cite{DoscherKeyl}.  That quantum coin tossing protocols offer
some advantage over classical ones was realized by Aharonov et al.\
\cite{Aharonov&2}, who introduced a protocol achieving a bias of
$\frac{1}{2\sqrt{2}}$ \cite{Aharonov&2,Spekkens&Rudolph2}.  For strong
coin tossing, it has been shown by Kitaev that in any protocol, at
least one party can achieve a bias greater than
$\frac{1}{\sqrt{2}}-\frac{1}{2}$ \cite{Kitaev}. It is not known
whether this figure represents an achievable bias.  The best known
bias to date (which is realized via a qutrit-based bit commitment
scheme) is $\frac{1}{4}$ \cite{Ambainis} and this is optimal for a
large set of protocols \cite{Spekkens&Rudolph}.  For weak coin
tossing, Kitaev's bound is known not to apply and lower biases than
$\frac{1}{\sqrt{2}}-\frac{1}{2}$ have been achieved (see for example
\cite{Mochon2} for the best bias to date).  Moreover, Ambainis has
shown that a protocol with bias $\epsilon>0$ must have a number of
rounds that grows as $\Omega(\log\log\frac{1}{\epsilon})$
\cite{Ambainis}.  

In this paper, we consider only non-relativistic protocols (in which
communications between parties can be effectively taken to be
instantaneous).  It is known that using two separated sites and
exploiting the impossibility of superluminal signalling, ideal coin
tossing can be implemented with perfect security \cite{Kent_CTBC}.

\subsection{Definitions}
In a coin tossing protocol, two separated and mistrustful parties,
Alice and Bob, wish to generate a shared random bit.  We consider a
model in which they do not initially share any resources, but have
access to trusted laboratories containing trusted error-free apparatus
for creating and manipulating quantum states. In general, a protocol
for this task may be defined to include one or more security
parameters, which we denote $N_1,\ldots,N_r$.

If both parties are honest, a coin tossing protocol guarantees that
they are returned the same outcome, $b\in\{0,1\}$ where outcome $b$
occurs with probability $\frac{1}{2}+\zeta_b(N_1,\ldots,N_r)$, or
``abort'' which occurs with probability $\zeta_2(N_1,\ldots,N_r)$,
where for each $j\in\{0,1,2\}$, $\zeta_j(N_1,\ldots,N_r)\rightarrow 0$
as the $N_i\rightarrow\infty$.  The {\it bias} of the protocol towards
party $P\in\{A,B\}$ is denoted
$\epsilon_P=\max\left(\epsilon_P^0,\epsilon_P^1\right)$, where $P$ can
deviate from the protocol in such a way as to convince the other
(honest) party that the outcome is $b$ with probability at most
$\frac{1}{2}+\epsilon_{P}^{b}+\delta_P^b(N_1,\ldots,N_r)$, where the
$\delta_P^b(N_1,\ldots,N_r)\rightarrow 0$ as the
$N_i\rightarrow\infty$.  We make no requirements of the protocol in
the case where both parties cheat.

The {\it bias} of the protocol is defined to be
$\max(\epsilon_A,\epsilon_B)$.  A protocol is said to be {\it
balanced} if $\epsilon_A^{b}=\epsilon_B^{b}$, for $b=0$ and $b=1$.

We define the following types of coin tossing:

{\bf Ideal Coin Tossing:}\qquad A coin tossing protocol is ideal if it
has $\epsilon_A=\epsilon_B=0$, that is, no matter what one party does
to try to bias the outcome, their probability of successfully doing so
is strictly zero.  It is then said to be {\it perfectly secure} if for
some finite values of $N_1,\ldots,N_r$, the quantities
$\zeta_j(N_1,\ldots,N_r)$ and $\delta_P^b(N_1,\ldots,N_r)$ are
strictly zero, and otherwise is said to be {\it secure}.

{\bf Strong Coin Tossing:}\qquad A strong coin tossing protocol is
parameterized by a bias, $\gamma$.  The protocol has the property that
$\epsilon_P^b\leq\gamma$ for all $P\in\{A,B\}$ and $b\in\{0,1\}$, with
equality for some $P$, $b$.

{\bf Weak Coin Tossing:}\qquad A weak coin tossing protocol is also
parameterized by a bias, $\gamma$.  It has the property that
$\epsilon_A^0\leq\gamma$ and $\epsilon_B^1\leq\gamma$, with equality
in one of the two inequalities.

\smallskip
In the next section we give a new protocol for strong coin tossing and
show that it is balanced and has bias $\frac{1}{4}$.

\section{The Protocol}
\begin{enumerate}
\item Alice creates $2$ copies of the state
   $\ket{\psi}=\frac{1}{\sqrt{2}}(\ket{00}+\ket{11})$ and sends the
   second qubit of each to Bob.
\item \label{Bch} Bob randomly selects one of the states to be used
  for the coin toss.  He informs Alice of his choice.
\item Alice and Bob measure their halves of the chosen state in the
  $\{\ket{0},\ket{1}\}$ basis to generate the result of the coin toss.
\item Alice sends her half of the other state to Bob who tests whether
  it is the state it should be by measuring the projection onto
  $\ket{\psi}$.  If his test fails, Bob aborts.
\end{enumerate}

\subsection{Alice's Bias}
Assume Bob is honest.  We will determine the maximum probability that
Alice can achieve outcome 0, $p_A$ (an analogous result follows by
symmetry for the case that Alice wants to bias towards 1).  Alice's
most general strategy is as follows. She can create a state in an
arbitrarily large Hilbert space,
$\ket{\Psi}\in\mathcal{H}_A\otimes\mathcal{H}_{A_1}\otimes\mathcal{H}_{B_1}\otimes\mathcal{H}_{A_2}\otimes\mathcal{H}_{B_2}$,
where $\mathcal{H}_A$ represents the space of an ancillary system
Alice keeps, $\mathcal{H}_{B_1}$ and $\mathcal{H}_{B_2}$ are qubit
spaces sent to Bob in the first step of the protocol, and
$\mathcal{H}_{A_1}$ and $\mathcal{H}_{A_2}$ are qubit spaces, one of
which will be sent to Bob for verification.  On receiving Bob's choice
of state in step \ref{Bch}, Alice can do one of two local operations
on the states in her possession, before sending Bob the relevant qubit
for verification.  Alice should choose her state and local operations
so as to maximize the probability that Bob obtains outcome $0$ and
does not detect her cheating.

Let us denote the state of the entire system by
\begin{equation}
\ket{\Psi}=a_{00}\ket{\phi_{00}}_{AA_1A_2}\ket{00}_{B_1B_2}+a_{01}\ket{\phi_{01}}_{AA_1A_2}\ket{01}_{B_1B_2}+a_{10}\ket{\phi_{10}}_{AA_1A_2}\ket{10}_{B_1B_2}+a_{11}\ket{\phi_{11}}_{AA_1A_2}\ket{11}_{B_1B_2},
\end{equation}
where $\{\ket{\phi_{ij}}_{AA_1A_2}\}_{i,j}$ are normalized states in
Alice's possession, and $\{a_{ij}\}_{i,j}$ are coefficients.  Suppose
Bob announces that he will use the first state for the coin toss.
There is nothing Alice can subsequently do to affect the probability
of Bob measuring 0 on the qubit in $\mathcal{H}_{B_1}$.  We can assume
that Bob makes the measurement on this qubit immediately on making his
choice.  Let us also assume that Alice discovers the outcome of this
measurement so that she knows the pure state of the entire system (we
could add a step in the protocol where Bob tells her, for
example\footnote{Such a step can only make it easier for Alice to
cheat, so security under this weakened protocol implies security under
the original one.}).  If Bob gets outcome $1$, then Alice cannot win.
On the other hand, if Bob gets outcome $0$, the state of the remaining
system becomes
\begin{equation}
\frac{a_{00}}{\sqrt{a_{00}^2+a_{01}^2}}\ket{\phi_{00}}_{AA_1A_2}\ket{0}_{B_2}+\frac{a_{01}}{\sqrt{a_{00}^2+a_{01}^2}}\ket{\phi_{01}}_{AA_1A_2}\ket{1}_{B_2},
\end{equation}
and Alice can win if she can pass Bob's test in the final step of the
protocol.  Since entanglement cannot be increased by local operations,
the system Alice sends to Bob in this case can be no more entangled
than this state.  Alice therefore cannot fool Bob into thinking she
was honest with probability greater than
$\frac{(a_{00}+a_{01})^2}{2(a_{00}^2+a_{01}^2)}$.  Using a similar
argument for the case that Bob chooses the second state for the coin
toss shows that Alice's overall success probability is at most
$\frac{1}{4}\left(2a_{00}^2+2a_{00}a_{01}+2a_{00}a_{10}+a_{01}^2+a_{10}^2\right)$.
Maximizing this subject to the normalization condition gives a maximum
of $\frac{3}{4}$, hence we have the bound
$p_A\leq\frac{3}{4}$. Equality is achievable within the original
protocol (i.e., without the additional step we introduced) by having
Alice use the state
\begin{eqnarray}
\sqrt{\frac{2}{3}}\ket{0000}_{A_1B_1A_2B_2}+\frac{1}{\sqrt{6}}\left(\ket{0011}_{A_1B_1A_2B_2}+\ket{1100}_{A_1B_1A_2B_2}\right),
\end{eqnarray}
and simply sending $\mathcal{H}_{A_1}$ or $\mathcal{H}_{A_2}$ to Bob
in the final step, depending on Bob's choice.  

The protocol is cheat-sensitive towards Alice---{\em any} strategy
which increases her probability of obtaining one outcome gives her a
non-zero probability of being detected.

\subsection{Bob's Bias}
Assume Alice is honest.  We will determine the maximum probability
that Bob can achieve the outcome 0, $p_B$.  The maximum probability
for outcome 1 follows by symmetry.  Bob seeks to take the qubits he
receives, perform some local operation on them, and then announce one
of them to be the coin-toss state such that the probability that Alice
measures 0 on her part of the state he announces is maximized.

Suppose that we have found the local operation maximizing Bob's
probability of convincing Alice that the outcome is 0.  Having
performed this operation and sent the announcement to Alice, the
outcome probabilities for Alice's subsequent measurement on the state
selected by Bob in the $\{\ket{0},\ket{1}\}$ basis are fixed.  Bob's
probability of winning depends only on this.  It is therefore
unaffected by anything Alice does to the other qubit, and, in
particular, is unaffected if Alice measures both of her qubits in the
$\{\ket{0},\ket{1}\}$ basis before looking at Bob's choice.  Such a
measurement commutes with Bob's local operation, so could be done by
Alice prior to Bob's operation without changing any outcome
probabilities.  If Alice does this measurement she gets outcome 1 on
both qubits with probability $\frac{1}{4}$.  In such a case, Bob
cannot convince Alice that the outcome is 0.  Therefore, we have
bounded Bob's maximum probability of winning via $p_B\leq\frac{3}{4}$.

To achieve equality, Bob can measure each qubit he receives in the
$\{\ket{0},\ket{1}\}$ basis, and if he gets one with outcome 0, choose
this state as the one to use for the coin toss.  There is no cheat
sensitivity towards Bob; he can use this strategy without fear of
being caught.

\section{Discussion}
We have introduced a new protocol for strong coin tossing which
achieves a bias of $\frac{1}{4}$.  Whilst this bias is not an
improvement over existing protocols, our protocol uses a conceptually
different approach to previous ones.  Rather than being built on
bit-commitment, the protocol works by attempting to share entanglement
between two parties, and then exploiting the resulting quantum
correlations to implement a coin toss.  This provides a further
illustration of the power of entanglement as a resource.  Furthermore,
our protocol requires only qubits for its implementation, whereas
bit-commitment based protocols cannot achieve such a bias without
using higher dimensional systems \cite{Spekkens&Rudolph}.

\bigskip
{\bf Additional Notes:} The protocol we present is a special case of a
protocol for random bit string generation found in
\cite{Kent_largeNQC}, where only one bit is sought\footnote{We thank
the anonymous referee for pointing this out.}.  However, the security analysis
in \cite{Kent_largeNQC} does not extend to the single bit case.  We
have also learned that this protocol was independently discovered by
Louis Salvail \cite{Rudolph_PC} whose work on this was not published.

\acknowledgments 

The author is grateful to Matthias Christandl and Adrian Kent for
useful discussions, and acknowledges financial support from EPSRC and
Trinity College, Cambridge.


\begin{thebibliography}{12}
\expandafter\ifx\csname natexlab\endcsname\relax\def\natexlab#1{#1}\fi
\expandafter\ifx\csname bibnamefont\endcsname\relax
  \def\bibnamefont#1{#1}\fi
\expandafter\ifx\csname bibfnamefont\endcsname\relax
  \def\bibfnamefont#1{#1}\fi
\expandafter\ifx\csname citenamefont\endcsname\relax
  \def\citenamefont#1{#1}\fi
\expandafter\ifx\csname url\endcsname\relax
  \def\url#1{\texttt{#1}}\fi
\expandafter\ifx\csname urlprefix\endcsname\relax\def\urlprefix{URL }\fi
\providecommand{\bibinfo}[2]{#2}
\providecommand{\eprint}[2][]{\url{#2}}

\bibitem[{\citenamefont{Blum}(1981)}]{Blum}
\bibinfo{author}{\bibfnamefont{M.}~\bibnamefont{Blum}}, in
  \emph{\bibinfo{booktitle}{CRYPTO}} (\bibinfo{year}{1981}), pp.
  \bibinfo{pages}{11--15}.

\bibitem[{\citenamefont{D\"oscher and Keyl}(2002)}]{DoscherKeyl}
\bibinfo{author}{\bibfnamefont{C.}~\bibnamefont{D\"oscher}} \bibnamefont{and}
  \bibinfo{author}{\bibfnamefont{M.}~\bibnamefont{Keyl}},
  \emph{\bibinfo{title}{An introduction to quantum coin tossing}},
  \bibinfo{howpublished}{e-print quant-ph/0206088} (\bibinfo{year}{2002}).

\bibitem[{\citenamefont{Aharonov et~al.}(2000)\citenamefont{Aharonov, Ta-Shma,
  Vazirani, and Yao}}]{Aharonov&2}
\bibinfo{author}{\bibfnamefont{D.}~\bibnamefont{Aharonov}},
  \bibinfo{author}{\bibfnamefont{A.}~\bibnamefont{Ta-Shma}},
  \bibinfo{author}{\bibfnamefont{U.~V.} \bibnamefont{Vazirani}},
  \bibnamefont{and} \bibinfo{author}{\bibfnamefont{A.~C.} \bibnamefont{Yao}},
  in \emph{\bibinfo{booktitle}{Proceedings of the 32nd annual ACM symposium on
  Theory of computing (STOC-00)}} (\bibinfo{publisher}{ACM Press},
  \bibinfo{address}{New York, NY, USA}, \bibinfo{year}{2000}), pp.
  \bibinfo{pages}{705--714}.

\bibitem[{\citenamefont{Spekkens and
  Rudolph}(2001{\natexlab{a}})}]{Spekkens&Rudolph2}
\bibinfo{author}{\bibfnamefont{R.~W.} \bibnamefont{Spekkens}} \bibnamefont{and}
  \bibinfo{author}{\bibfnamefont{T.}~\bibnamefont{Rudolph}},
  \bibinfo{journal}{Quantum Information and Computation}
  \textbf{\bibinfo{volume}{2}}, \bibinfo{pages}{66}
  (\bibinfo{year}{2001}{\natexlab{a}}).

\bibitem[{\citenamefont{Kitaev}()}]{Kitaev}
\bibinfo{author}{\bibfnamefont{A.}~\bibnamefont{Kitaev}},
  \bibinfo{note}{(unpublished), proof recreated in \cite{Ambainis&}}.

\bibitem[{\citenamefont{Ambainis}(2004)}]{Ambainis}
\bibinfo{author}{\bibfnamefont{A.}~\bibnamefont{Ambainis}},
  \bibinfo{journal}{Journal of Computer and System Sciences}
  \textbf{\bibinfo{volume}{68}}, \bibinfo{pages}{398} (\bibinfo{year}{2004}),
  ISSN \bibinfo{issn}{0022-0000}.

\bibitem[{\citenamefont{Spekkens and
  Rudolph}(2001{\natexlab{b}})}]{Spekkens&Rudolph}
\bibinfo{author}{\bibfnamefont{R.~W.} \bibnamefont{Spekkens}} \bibnamefont{and}
  \bibinfo{author}{\bibfnamefont{T.}~\bibnamefont{Rudolph}},
  \bibinfo{journal}{Physical Review A} \textbf{\bibinfo{volume}{65}},
  \bibinfo{pages}{012310} (\bibinfo{year}{2001}{\natexlab{b}}).

\bibitem[{\citenamefont{Mochon}(2005)}]{Mochon2}
\bibinfo{author}{\bibfnamefont{C.}~\bibnamefont{Mochon}},
  \bibinfo{journal}{Physical Review A} \textbf{\bibinfo{volume}{72}},
  \bibinfo{pages}{022341} (\bibinfo{year}{2005}).

\bibitem[{\citenamefont{Kent}(1999)}]{Kent_CTBC}
\bibinfo{author}{\bibfnamefont{A.}~\bibnamefont{Kent}},
  \bibinfo{journal}{Physical Review Letters} \textbf{\bibinfo{volume}{83}},
  \bibinfo{pages}{5382} (\bibinfo{year}{1999}).

\bibitem[{\citenamefont{Kent}(2002)}]{Kent_largeNQC}
\bibinfo{author}{\bibfnamefont{A.}~\bibnamefont{Kent}}, in
  \emph{\bibinfo{booktitle}{Proceedings of the 6th International Conference on
  Quantum Communication, Measurement and Computing (QCMC'02)}}, edited by
  \bibinfo{editor}{\bibfnamefont{J.~H.} \bibnamefont{Shapiro}}
  \bibnamefont{and} \bibinfo{editor}{\bibfnamefont{O.}~\bibnamefont{Hirota}}
  (\bibinfo{publisher}{Rinton Press Inc.}, \bibinfo{year}{2002}),
  \bibinfo{note}{also available at http://arxiv.org/abs/quant-ph/0212043}.

\bibitem[{\citenamefont{Rudolph}()}]{Rudolph_PC}
\bibinfo{author}{\bibfnamefont{T.}~\bibnamefont{Rudolph}},
  \bibinfo{howpublished}{Personal Communication}.

\bibitem[{\citenamefont{Ambainis et~al.}(2004)\citenamefont{Ambainis, Buhrman,
  Dodis, and R\"ohrig}}]{Ambainis&}
\bibinfo{author}{\bibfnamefont{A.}~\bibnamefont{Ambainis}},
  \bibinfo{author}{\bibfnamefont{H.}~\bibnamefont{Buhrman}},
  \bibinfo{author}{\bibfnamefont{Y.}~\bibnamefont{Dodis}}, \bibnamefont{and}
  \bibinfo{author}{\bibfnamefont{H.}~\bibnamefont{R\"ohrig}}, in
  \emph{\bibinfo{booktitle}{Proceedings of the 19th IEEE Annual Conference on
  Complexity}} (\bibinfo{publisher}{IEEE Computer Society},
  \bibinfo{year}{2004}), pp. \bibinfo{pages}{250--259}, ISBN
  \bibinfo{isbn}{0-7695-2120-7}.

\end{thebibliography}
\end{document}